\documentclass[preprint,fleqn,showpacs,showkeys]{revtex4} 
\usepackage{graphicx}
\usepackage{epstopdf}
\usepackage{amssymb}
\usepackage{amsmath}
\usepackage{bm}
\usepackage[font=footnotesize]{subfig}
\begin{document}
\setcounter{page}{0}
\title[]{Design Study of a Superconducting Gantry for Carbon Beam Therapy}
\author{J. \surname{Kim}}
\author{M. \surname{Yoon}}
\email{moohyun@postech.ac.kr}
\thanks{Fax: +82-54-279-3099 }
\affiliation{Department of Physics,  Pohang University of Science and Technology, 
77 Cheongam-Ro, Nam-Gu, Pohang, Gyeongbuk, Korea 790-784 }

\date[]{Received November 2015}

\begin{abstract}
This paper describes the design study of a gantry for a carbon beam.
The designed gantry is compact such that its size is comparable to the size of the
proton gantry. This is possible by introducing superconducting double helical
coils for dipole magnets.
The gantry optics is designed in such a way that it provides
rotation-invariant optics and variable beam size  as well as point-to-parallel scanning of a beam.
For large-aperture magnet, three-dimensional magnetic field distribution is obtained
by invoking a computer code, and a number of particles are
tracked by integrating equations of motion numerically together with three-dimensional
interpolation.
The beam-shape distortion due to the fringe field
is reduced to an acceptable level by optimizing the coil windings
with the help of genetic algorithm.
Higher-order transfer coefficients are calculated and shown to
be reduced greatly with appropriate optimization of the coil windings.
\end{abstract}



\maketitle

\section{INTRODUCTION}
Treatment of malignant tumor by means of the ion-beam therapy (or 
equivalently ion-beam radiotherapy)
has become increasingly popularized in recent decades. In the ion-beam 
therapy, an accelerated beam penetrates into the human body and deposits most of 
its energy into the DNA of tumor cells while mitigating the damage of normal tissues 
surrounding the tumor (Bragg peak). Compared to photons (or x-rays)
ions have better dose-depth distribution.

Usually protons are used in the ion-beam therapy. At present about
50 proton-therapy facilities are in operation around the world and more than
105000 patients have been treated so far. But 
because of better physical and biological effect, heavier ions 
(usually $^{12}$C$^{6+}$ ions) are also employed. There are 
eight such facilities (six in Japan and two in Europe) active currently.

To deliver an ion beam from any direction to the tumor inside a patient, many 
ion-beam therapy facilities use gantries which are rotatable beam-transport systems
mainly equipped with dipole magnets, quadrupoles, and scanning magnets togther with various 
diagnostic instruments. 
The gantry is in general massive and voluminous and so it is expensive to build. For example, 
the second proton gantry at Paul Scherrer Institut (PSI) in Switzerland is about
12 m wide and 4 m high and it's weight is approximately 100 tons. 

The magnetic rigidity of the carbon ion
is approximately 2.5 times higher than the proton, and so carbon ion requires higher energy 
than the proton energy does. For instance, 
for 30 cm penetration depth in water 430 MeV/u $^{12}$C$^{6+}$ is required 
whereas for proton only 220 MeV is needed. The rigidity of 430 MeV/u $^{12}$C$^{6+}$ is
6.623 T m. This implies that a gantry for carbon ion demands much larger space and heavier structure
than proton gantry. For example, 
the space occupied by the carbon gantry in the Heidelberg Ion Therapy (HIT) facility
is 22 m long and 14 m high~\cite{HIT} and the total weight of this gantry system is approximately
600 t which is twice larger and five times heavier than the proton gantry in general.
Obviously such a structure is not easy to build and operate. 

Therefore, it is an important task to reduce the size of the carbon gantry.
Various designs were reported  so far. 

In 2010, a preliminary study for design of a carbon gantry based on
superconducting toroidal combined-function magnet was reported~\cite{Yoon}.
This study confirmed that by employing 5 T superconducting magnets the size of the carbon
gantry could be reduced to proton gantry, approximately 10 m by 4 m.  The designed gantry is similar to
the HIT design retaining all the beam optical features such as variable beam size from 4 mm to
10 mm, point to parallel scanning capability, and rotational invariant optics. The scanning magnets
cover $\pm$10 cm region in horizontal and vertical planes. In addition, our study reveals that
the beam shape at the isocenter is significantly distorted  due to nonlinear fields in the
fringe field region of the large aperture 90$^\circ$ dipole magnet. Sextuple components in the
fringe field are found to be the cause of this problem.

To cure the beam shape distortion the coil winding in the main body of the dipole magnet
has been redistributed in such a way that the resulting sextuple components in the
main body compensate for the nonlinear field in the fringe field region. To find the optimized
coil winding the genetic algorithm (GA)~\cite{GA} has been invoked. Results indicated
the compensation was made appropriately and subsequent preliminary particle tracking studies
revealed to some extent the validity of the optimization of the GA.

This paper extends the previous studies and presents result of
the further optimization of the coil winding. In Section II,
we show a brief introduction of the beam optics of the designed gantry based on the
superconducting magnet.  This section is brief as mathematical result is already
described elsewhere~\cite{HIT}. Section III describes the 90$^\circ$ combined function
dipole magnet. Section IV presents the method and results of the particle tracking.
Finally, summary and conclusion is given in Section V.

\section{GANTRY BEAM OPTICS}
The designed gantry is of similar type to the HIT carbon gantry~\cite{HIT}. Figure 1
shows the layout of the gantry which shows six quadrupole magnets, two 45$^\circ$ dipole magnets,
one 90$^\circ$ dipole magnet, and two scanning magnets.

The six quadrupole magnets are used for beam size control at the isocenter in 4 - 10 mm range
(for example, $R_{12}$ = 4 mm and $R_{34}$ = 4 mm), dispersion-free optics at the isocenter
($R_{16}$ = $R_{26}$ = 0), and rotation invariant optics ($R_{11}$ = $R_{33}$ = 0). Here $R_{ij}$
is the 6$\times$6 linear transfer matrix~\cite{Brown}, ${\bf X} = R {\bf X}_0$ with ${\bf X}$ and ${\bf X}_0$ being
1 by 6 orbit vectors, $\bf X$ = $(x, x', y, y', l, \delta)^T$. The
matrix equation is explicitly given by
\begin{equation} \label{eq1}
 \left( \begin{array}{c}
x \\
x' \\
y \\
y' \\
l \\
\delta
             \end{array}  \right)  =
\left( \begin{array}{cccccc}
R_{11}  & R_{12} & R_{13} & R_{14} & R_{15} & R_{16} \\
R_{21}  & R_{22} & R_{23} & R_{24} & R_{25} & R_{26} \\
R_{31}  & R_{32} & R_{33} & R_{34} & R_{35} & R_{36} \\
R_{41}  & R_{42} & R_{43} & R_{44} & R_{45} & R_{46} \\
R_{51}  & R_{52} & R_{53} & R_{54} & R_{55} & R_{56} \\
R_{61}  & R_{62} & R_{63} & R_{64} & R_{65} & R_{66} \\
\end{array}  \right)
 \left( \begin{array}{c}
x_{0} \\
x'_{0} \\
y_{0} \\
y'_{0} \\
l_{0} \\
\delta_{0}
             \end{array}  \right )
\end{equation}

Figure 2 shows the beam envelops along the beam path length
for $\pm$90$^\circ$ in 10$^\circ$ step for 4 mm beam size at the iso center. It is seen that
beam size at the isocenter does not depend on the rotation angle of the gantry. Also dispersion
is shown to be localized between dipole magnets.

Upstream of the last 90$^\circ$ combined-function dipole magnet two scanning magnets are placed,
one for horizontal
scanning and other for vertical scanning respectively. These magnets steer the beam by $\pm$10 cm
in each plane. Locations of the scanning magnets were chosen to make point-to-parallel
optics (i. e. $R_{22}$ = 0
 from the position of the horizontal scanning magnet to the isocenter and
$R_{44}$ = 0   from the vertical scanning magnet to the isocenter) to minimize the
radiation power density of the patient's body. Thus in this design the source-to-axis distance
(SAD) is infinity.

\section{90$^\circ$ BENDING MAGNET}
In Figure 1 the last magnet located upstream of the isocenter is a 90$^\circ$ dipole magnet.
This magnet is the most critical element in a carbon gantry because of its large aperture.
The $90^{\circ}$ bending magnet requires combined magnetic multipole components
 in a body to satisfy the beam optics condition of the designed gantry and to correct
 high order effects.

Previously a novel winding concept was introduced for this magnet~\cite{Meinke,Caspi1,Caspi2,Yoon}. Superposing
two solenoid-like coils oppositely skewed with respect to a cylindrical axis $\cos\theta$ distribution
of current density is achieved and the resulting magnetic field in the bore is a dipole.
Coils wraps quater of torus shape and the beam is transported inside
the wraped coils. The magnet can create not only dipole and quadrupole components but also sextupole
and higher order multipoles. The winding follows the equation which
is a function of the toroidal angle $\varphi $ and poloidal angle $\theta$ in toroidal coordinates
(Figure 3).
The winding equation for torus is given by
 \begin{equation}
 \varphi = \frac{\theta }{n} + a_{0}\sin\theta + a_{1}\sin2\theta + a_{2}\sin3\theta + \cdots
 \end{equation}
where $n$ is the number of windings in a full torus and $a_{0}$, $a_{1}$, $a_{2}$, ...
 represent coefficients of  multipoles. Specifically, $a_{0}$ controls dipole field strength,
 $a_{1}$ controls quadrupole field strength, $a_{2}$ controls sextupole field strength, and so on.

Figure 4 shows three-dimensional view of the designed $90^{\circ}$ bending magnet.
Strength of the magnetic dipole field is chosen to be 5.22 T at the center
which makes 430 MeV carbon ion be centered at 1.269 m in radius. Quadrupole field is
set to have the field index of 0.5 to yield equal focussing in both planes but after design
the position of the scanning magnets are readjusted from the calculation of the
linear transfer matrix based on the multi-particle tracking.

Magnetic fields of the $90^{\circ}$ magnet was obtained by invoking a three-dimensional
magnetic field solver OPERA-3D.  Figure 5 shows the dipole field strength along the beam path
of the $90^{\circ}$ bending magnet.

\section{PARTICLE TRACKING}
To verify the validity of the gantry design, particle tracking has been performed.
For input beam parameters, 2 mm $\times$ 0.5 mrad horizontally, 10 mm $\times$ 0.5 mrad
vertically, and momentum spread 0.2$\%$ were chosen, which is identical to the GSI design~\cite{HIT}.
Total 1000 particle trajectories were obtained by numerically integrating the equations
of motion in cylindrical coordinate system. During the integration, magnetic fields were
obtained by interpolating the fields in the three dimensional space.

First, the reference particle was searched by tracing a particle at the center of the
90$^\circ$ dipole magnet and particle energy was adjusted so that the resulting reference
particle exits the magnet at 45$^\circ$. Once the reference particle was found,
transfer coefficients were obtained by tracking specially arranged particles. For integrating
the equations of motion, the
fourth-order Runge-Kutta integration method has been used.
Positions of the scanning magnets were determined so as to yield
the point-to-parallel optics. Then 1000 particles distributed as described in the above
were tracked from the starting position at the entrance of the gantry to the
position of the scanning magnets. At the scanning magnets kicks were applied in uniform steps
and the particles were traced to the isocenter. In the
calculation, the following parameters have been used: $R$ = 1.269 m, $\rho$ = 0.186 m,
$n$ = 864, $a_0$ = 0.162, $a_1$ = -0.00877018, $a_2$ = -0.00098121, and the
current density $J$ = 34287 A/m$^2$ with the coil radius of 0.5 mm.

Figure 6 shows the kicks applied at the scanning magnets to span $\pm$10 cm. Figure 7
shows the resulting beam distribution at the isocenter. It is seen that beam shapes
are distorted significantly. Although kicks are applied in equal steps, resulting
positions of the beam center at large kicks deviate from the linearity too much.
Main cause of the shape distortions and the nonlinearity is found to be due to
large sextupole components in the fringe field region of the 90$^\circ$ magnet.

To cure the beam shape distortion, the coefficients $a_i$ have been readjusted
by utilizing the genetic algorithm~\cite{GA}.
Biot-Savart law was used for calculation of the magnetic field due to coil windings.
Final coil windings were then
used for new magnetic field calculation by OPERA 3D code. Repeating the same
procedure described above, the improved beam distribution at the isocenter is given in
Figure 8. Significant improvement is achieved as this figure indicates.
New multipole coefficients are $a_0$ = 0.162 , $a_1$ = -0.00877018, and
$a_2$ = 0.000371945.

For quantitative comparison, transfer coefficients were calculated~\cite{Yoon2}.
The transfer coefficients are defined as
\begin{eqnarray}\label{eq3}
X_i &=& \sum_{j=1}^6 R_{ij}X_{j0}+ \sum_{\substack {j,k \\ j \le k}}T_{ijk}X_{j0}X_{k0}+ \nonumber \\
& & \quad\quad
\sum_{\substack{j,k,l \\ j \le k \le l}}U_{ijkl}X_{j0}X_{k0}X_{l0} + \nonumber \\
& & \quad\quad
\sum_{\substack{j,k,l,m\\ j\le k \le l \le m}}W_{ijklm}X_{j0}X_{k0}X_{l0}X_{m0} + \cdots
\end{eqnarray}
where $R$, $T$, $U$, and $W$ are first-, second-, third-, and fourth-order transfer coefficients, respectively.
$X_{i0}$ and $X_i$  are the $i^{th}$ phase-space coordinates at the input and
output positions, respectively.

Table I shows the comparison of the second-order transfer coefficients $T_{ijk}$
before and after optimization of the coil windings. It is seen that
after adjustment of the coil winding the second-order coefficients
have been reduced significantly.

\begin{table}
\caption{ Comparison of $T_{ijk}$ ( cm/mrad$^{2}$ , mrad/mrad$^{2}$) before and after the optimization of the coil windings}
\begin{ruledtabular}
\begin{tabular}{cccccc}
$T_{ijk}$  & Before improvement & After improvement \\
\hline
\colrule
$T_{122}$ & -3.263 & 1.627 $\times$  10$^{-5}$ \\
$T_{144}$ & -3.249 & -1.817 $\times$  10$^{-4}$ \\
$T_{222}$ & -10.08 & 1.122 $\times$  10$^{-4}$ \\
$T_{244}$ & -10.09 & -5.026 $\times$  10$^{-4}$ \\
$T_{322}$ & -18.61 & 4.557 $\times$  10$^{-8}$ \\
$T_{344}$ & -18.61 & 1.170 $\times$  10$^{-7}$ \\
$T_{422}$ & -2.793 & 1.623 $\times$  10$^{-7}$ \\
$T_{444}$ & -2.796 & 4.524 $\times$  10$^{-7}$ \\
$T_{124}$ & 3.246 & 5.444 $\times$  10$^{-8}$ \\
$T_{224}$ & 10.09 & 3.983 $\times$  10$^{-7}$ \\
$T_{324}$ & 18.61 & -1.008 $\times$  10$^{-4}$ \\
$T_{424}$ & 2.787 & -1.098 $\times$  10$^{-3}$ \\
\end{tabular}
\end{ruledtabular}
\label{table1}
\end{table}

\section{CONCLUSIONS}
Beam optical studies for a  gantry for carbon ion transport have been performed.
The designed gantry is compact in the sense that its size is comparable to the size of
existing proton gantries. Superconducting double helical coils were used for 90$^\circ$ bending magnet.
For this magnet, magnetic fields were calculated with the three-dimensional code and
particles trajectories were calculated numerically by integrating equations of motion.
We have found that there was a beam shape distortion at the isocenter which is large when the beam
is steered at large angle. The distortion was found to be due to nonlinear fields in the fringe field region
of the bending magnet of large aperture. We were able to demonstrate that the nonlinear distortion
could be cured by means of adjusting coil winding distribution.

Our study needs to be extended by including an iron wrapped around the coils. The iron can shield the stray magnetic
field as well as enhancing the magnetic field inside the bore. Design study for this case is in progress.

\begin{acknowledgments}
This work is supported by POSTECH Basic Science Research
Institute Grant and the Heavy Ion Accelerator Project at Korea Institute
of Radiological and Medical Science (KIRAMS).
\end{acknowledgments}

\begin{figure}
\centering
\includegraphics[width=7cm]{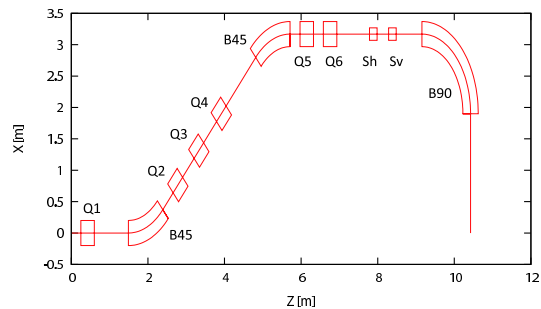}
\caption{Schematic layout for carbon gantry. Q1 $\sim~$Q6 are quadrupole magnets.
B45 is a 45$^\circ$ dipole magnet. Sh, Sv are horizontal steering magnet and
vertical steering magnet respectively. B90 is a 90$^\circ$ combined-function dipole magnet.} \label{fig.1}
\end{figure}

\begin{figure}
\centering
\includegraphics[width=7cm]{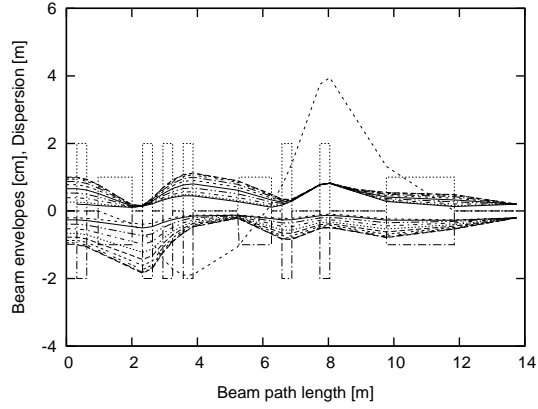}
\caption{Beam size variation along the beam path for $\pm$90$^\circ$ in 10$^\circ$ step. Dashed lines
converging to a single point at the gantry end indicate beam envelopes for each rotation step. A single dashed line indicates 
horizontal dispersion function along the gantry.} \label{fig.2}
\end{figure}

\begin{figure}
\centering
\includegraphics[width=7cm]{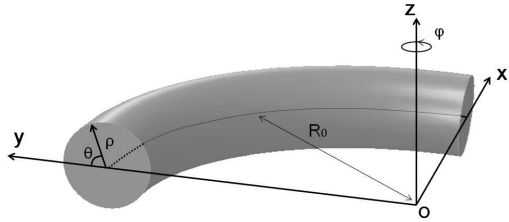}
\caption{Toroidal coordinates and cartesian coordinates:
$R_{0}$ is the major radius of torus and
$\rho$ is the minor radius of torus. $\varphi $is the toroidal angle. $\theta$ is the poloidal angle. $O$ is
the origin of the torus. } \label{fig.3}
\end{figure}

\begin{figure}
\centering
\includegraphics[width=7cm]{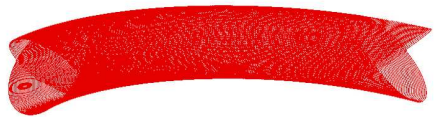}
\caption{$90^{\circ}$ bending magnet with double helical winding} \label{fig.4}
\end{figure}

\begin{figure}
\includegraphics[width=7cm]{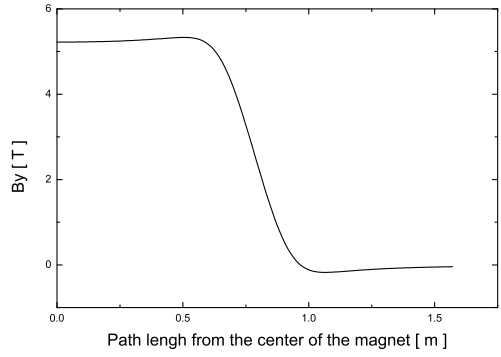}
\caption{Dipole field along the path length from the center of the $90^{\circ}$ bending magnet.}
\label{fig5}
\end{figure}

\begin{figure}
\includegraphics[width=7cm]{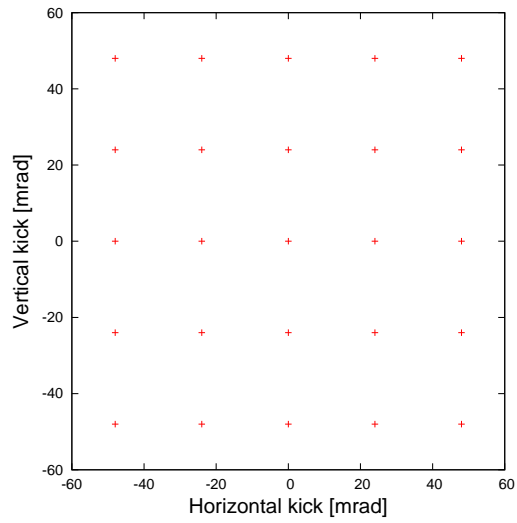}
\caption{Distribution of kick angles at horizontal and vertical scanning magnets }
\label{fig6}
\end{figure}

\begin{figure}
\includegraphics[width=7cm]{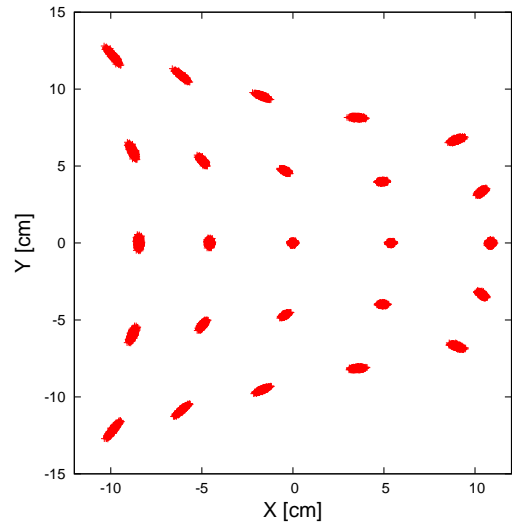}
\caption{Beam spots at the isocenter before optimization}
\label{fig7}
\end{figure}

\begin{figure}
\includegraphics[width=7cm]{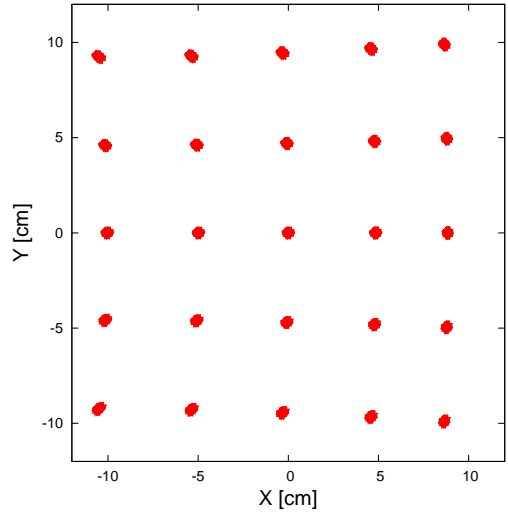}
\caption{Beam spots at the isocenter after readjusting coil winding distribution}
\label{fig8}
\end{figure}

\end{document}